\documentclass[pra,aps,amsmath,amssymb,showkeys,showpacs]{revtex4}
\usepackage{graphicx}
\newcommand{\pen}{\openone}

\newcommand{\ca}{{\mathcal{C}}}
\newcommand{\mc}{{\mathcal{M}}}
\newcommand{\tca}{{\mathcal{R}}}
\newcommand{\pbr}{{\mathrm{Pr}}}
\newcommand{\msa}{\mathsf{A}}
\newcommand{\msb}{\mathsf{B}}
\newcommand{\va}{\vec{a}}
\newcommand{\vb}{\vec{b}}
\newcommand{\bsg}{\boldsymbol{\sigma}}

\begin{document}
\clearpage
\preprint{}

\title{Tests for quantum contextuality in terms of $q$-entropies}
\author{Alexey E. Rastegin}
\email{rast@api.isu.ru; alexrastegin@mail.ru}
\affiliation{Department of Theoretical Physics, Irkutsk State University,
Gagarin Bv. 20, Irkutsk 664003, Russia}

\begin{abstract} The information-theoretic
approach to Bell's theorem is developed with use of the
conditional $q$-entropies. The $q$-entropic measures fulfill many
similar properties to the standard Shannon entropy. In general,
both the locality and noncontextuality notions are usually treated
with use of the so-called marginal scenarios. These hypotheses
lead to the existence of a joint probability distribution, which
marginalizes to all particular ones. Assuming the existence of
such a joint probability distribution, we derive the family of
inequalities of Bell's type in terms of conditional $q$-entropies
for all $q\geq1$. Quantum violations of the new inequalities are
exemplified within the Clauser--Horne--Shimony--Holt (CHSH) and
Klyachko--Can--Binicio\v{g}lu--Shumovsky (KCBS) scenarios. An
extension to the case of $n$-cycle scenario is briefly mentioned.
The new inequalities with conditional $q$-entropies allow to
expand a class of probability distributions, for which the
nonlocality or contextuality can be detected within entropic
formulation. The $q$-entropic inequalities can also be useful in
analyzing cases with detection inefficiencies. Using two models of
such a kind, we consider some potential advantages of the
$q$-entropic formulation.
\end{abstract}


\maketitle

\pagenumbering{arabic}
\setcounter{page}{1}

\section{Introduction}\label{sec1}

The notion of entanglement plays a key
role in studies of non-classical features of quantum theory. Due
to impressive advances, entangled quantum states are now treated
as tools for information processing \cite{nielsen}. An existence
of purely quantum correlations was emphasized in the
Schr\"{o}dinger ``cat paradox'' paper \cite{cat35} and the
Einstein--Podolsky--Rosen paper \cite{epr35}. Nonlocal
correlations are brightly manifested in specified experiments
similar to Bohm's version of the EPR argument \cite{bohm51}. In
such experiments, spacelike separated observers share subsystems
of an entangled quantum system. From an intuitive viewpoint, the
following assumptions seem to be relevant. First, one assumes that
physical quantities have well established values previous to any
measurement. Second, no signals can travel faster than the speed
of light. A less known point is the assumption of measurement
independence \cite{hall10,bgis11}. The assumptions lead to
restrictions commonly referred to as Bell inequalities. The
fundamental result is that such restrictions on correlations are
overcome within quantum mechanics \cite{bell64}. Today, a role of
Bell inequalities widely ranges from the foundations \cite{az99}
up to applications in quantum information processing like quantum
key distillation \cite{agms04,ags03} and randomness expansion
\cite{panat10}. In a certain sense, Leggett--Garg inequalities
\cite{carg85} are closely related to Bell ones. On the other hand,
Leggett--Garg inequalities probe correlations of a single system
measured at different times. A theoretical background,
experimental tests and some proposals for such inequalities are
reviewed in Ref. \cite{eln13}. In Ref. \cite{uksr13}, this issue
is examined within the entropic approach.

Like the locality, the noncontextuality assumption is also natural
from the classical viewpoint. In quantum theory, this pertains
only to  mutually compatible observables, which are simultaneously
diagonalizable. Hence, performed measurement of one of such
observables does not stipulate results of further measurements of
other. It turns out that no noncontextual hidden-variable models
can reproduce all the predictions of quantum theory
\cite{bell66,ks67}. This result known as the Kochen--Specker
theorem was independently obtained by Bell (for details, see Ref.
\cite{mermin93}). The recent paper \cite{oresh12} focused on the
causality, which is also deeply rooted in our understanding of the
macro world. In quantum mechanics, we may conceive
situations in which a single event can be equally a cause and an
effect of another one \cite{oresh12}. As discussed results concern
measurement statistics, they are statements about probability
distributions. In general, there are various ways to express
probabilistic properties. Although many formulations of Bell's
theorem use inequalities, the Greenberger--Horne--Zeilinger
argument has provided a claim without inequalities \cite{ghsz90}.
The EPR and GHZ states can give suitable tools in considering
three-partite entanglement \cite{avc03}. Bell inequalities can be
treated geometrically within multilinear-contraction framework
\cite{scapw10}. Entropic formulations of Bell's theorem have been
proposed in Ref. \cite{BC88} and further examined in Ref.
\cite{cerf97}. Various entropic measures are indispensable tools
in analyzing secure protocols \cite{renner05}.

There exist several concrete scenarios to realize Bell's theorem
as an experimentally tested statement. The
Clauser--Horne--Shimony--Holt (CHSH) scenario \cite{chsh69} is
probably the most known setup of such a kind. The CHSH inequality
imposes a restriction on  mean values of the corresponding
observables. Its violation allows to renounce local
hidden-variable models \cite{agr82,adr82}. The
Klyachko--Can--Binicio\v{g}lu--Shumovsky (KCBS) scenario
\cite{kly08} is also the subject of active research. The entropic
approach has been applied to both the CHSH \cite{BC88,rchtf12} and
KCBS scenarios \cite{rchtf12,krk12}. These scenarios can be
treated respectively as the $n=4$ and $n=5$ cases of more general
$n$-cycle scenario \cite{lsw11,aqbcc12}. For the $n$-cycle
scenario, the quantum violations occur for all $n$, though
technical motives make their observation harder for large $n$
\cite{bcdfa13}. Various aspects of entropic inequalities for
marginal problems are considered in Ref. \cite{chfr13}. The
information-theoretical results are usually expressed in terms of
standard functionals based on the Shannon entropy. Applying
statistical methods in numerous topics, some extensions were found
to be useful. The R\'{e}nyi \cite{renyi61} and Tsallis
\cite{tsallis} entropies are both especially important
generalizations. The nonlocality and contextuality are genuine
quantum features related also to the field of quantum information
processing. So, it is of importance to develop the entropic
approach to Bell inequalities with use of generalized entropies.

The aim of the present paper is to study information-theoretic
formulations of Bell's theorem in terms of the conditional Tsallis
entropies. It turns out that important achievements can be reached
in this way. The paper is organized as follows. In Sect. 2, basic
properties of the Tsallis entropies are recalled. We also prove
two required statements about the conditional $q$-entropy, one of
them for $q\geq1$ only. In Sect. 3, marginal scenarios are
discussed from the viewpoint of their use in studying Bell
inequalities. For the CHSH scenario, inequalities of Bell's type
in terms of the conditional $q$-entropies are obtained in Sect. 4.
We also mention an extension to the $n$-cycle scenario, which is
currently the subject of active research \cite{lsw11,aqbcc12}. In
Sect. 5, we consider $q$-entropic inequalities with $q\geq1$ for
the KCBS scenario. In both the cases, violations of the obtained
inequalities could be tested in the experiment. As is shown,
$q$-entropic inequalities with suitably chosen $q>1$ can detect
the nonlocality or contextuality of some probability
distributions, for which inequalities with the standard entropy
fail. We also analyze the $q$-entropic inequalities within two
models of detection inefficiencies. In other words, the family of
$q$-entropic inequalities is much more powerful to reveal such
properties. In Sect. 6, we conclude the paper with a summary of
results.

\section{Conditional $q$-entropies and their properties}\label{sec2}

In this section, we recall definitions of the Tsallis
entropies and related conditional entropies. Required properties
of these entropic functionals are discussed as well. Let the
variable $A$ take values on the set $\Omega_{A}$ with
corresponding probability distribution
$\bigl\{p(a):{\>}a\in\Omega_{A}\bigr\}$. The Tsallis entropy of
order $q>0\neq1$ is defined by \cite{tsallis}
\begin{equation}
H_{q}(A):=\frac{1}{1-q}{\,}\left(\sum_{{\,}a\in\Omega_{A}} p(a)^{q}
- 1 \right)
{\,}. \label{tsaent}
\end{equation}
With the factor $\left(2^{1-q}-1\right)^{-1}$ instead of
$(1-q)^{-1}$, this entropic form was derived from several axioms
by Havrda and Charv\'{a}t \cite{havrda}. Let $B$ be another
variable taking values on the set $\Omega_{B}$ with probability
distribution $\bigl\{p(b):{\>}b\in\Omega_{B}\bigr\}$. The joint
$q$-entropy $H_{q}(A,B)$ is defined similarly to Eq.
(\ref{tsaent}), but with joint probabilities $p(a,b)$ instead of
$p(a)$. It is sometimes convenient to rewrite the entropy
(\ref{tsaent}) as
\begin{equation}
H_{q}(A)=-\sum_{a\in\Omega_{A}}p(a)^{q}\ln_{q}p(a)
=\sum_{a\in\Omega_{A}}p(a){\,}\ln_{q}\frac{1}{p(a)}
\ . \label{tsaln}
\end{equation}
The $q$-logarithm $\ln_{q}x=\bigl(x^{1-{q}}-1\bigr)/(1-{q})$
is defined for $q>0\not=1$ and $x>0$, and it obeys
$\ln_{q}(1/x)=-x^{q-1}\ln_{q}x$. In the limit $q\to1$, we obtain
$\ln_{q}x\to\ln{x}$ and the standard Shannon entropy
\begin{equation}
H_{1}(A)=-\sum\nolimits_{a\in\Omega_{A}}p(a){\,}\ln{p(a)}
\ . \label{shaln}
\end{equation}
For brevity, we will usually omit the symbol of the set
$\Omega_{A}$ in entropic sums. Properties of quantum counterpart
of the entropy (\ref{tsaent}) are examined in Ref. \cite{raggio}.
Applications of various entropic functions in studying quantum
systems are discussed in the book \cite{bengtsson}.

To analyze more realistic cases with detector inefficiencies, the
following questions will rise. For the given $\eta\in[0;1]$ and
probability distribution $\bigl\{p(a):{\>}a\in\Omega_{A}\bigr\}$,
the set
\begin{equation}
\{p_{\eta}\}:=\bigl\{\eta{p}(a):{\>}a\in\Omega_{A}\bigr\}\cup\{1-\eta\}
\label{peta}
\end{equation}
is a probability distribution as well. This probability
distribution corresponds to some random variable $A_{\eta}$. We
aim to relate the entropy $H_{q}(A_{\eta})$ with $H_{q}(A)$ and
the binary $q$-entropy
\begin{equation}
h_{q}(\eta):=-{\,}\eta^{q}\ln_{q}\eta-(1-\eta)^{q}\ln_{q}(1-\eta)
\ . \label{bend}
\end{equation}
From three probability distributions, we can built another
probability distribution
\begin{equation}
\{p_{\eta\eta}\}:=
\bigl\{\eta^{2}p(a)\bigr\}
\cup\bigl\{\eta(1-\eta)p(b)\bigr\}\cup\bigl\{\eta(1-\eta)p(c)\bigr\}\cup\bigl\{(1-\eta)^{2}\bigr\}
\ . \label{prete}
\end{equation}
In this case, we aim to relate the obtained $q$-entropy with the
$q$-entropies of the initial probability distributions. The
following statement takes place.

\newtheorem{t20}{Lemma}
\begin{t20}\label{lem0}
Let random variable $A_{\eta}$ take its values according to the
probability distribution (\ref{peta}). For all $q>0$, the
$q$-entropies satisfy
\begin{equation}
H_{q}(A_{\eta})=\eta^{q}H_{q}(A)+h_{q}(\eta)
\ . \label{rlem0}
\end{equation}
Let random variable $A_{\eta\eta}$ take its values according to
the probability distribution (\ref{prete}). For all $q>0$, the
$q$-entropies satisfy
\begin{equation}
H_{q}(A_{\eta\eta})=
\eta^{2q}H_{q}(A)+\eta^{q}(1-\eta)^{q}\bigl(H_{q}(B)+H_{q}(C)\bigr)+\bigl(\eta^{q}+(1-\eta)^{q}+1\bigr)h_{q}(\eta)
\ . \label{rlem00}
\end{equation}
\end{t20}

{\bf Proof.} We first assume $q>0\neq1$. Substituting the
distribution (\ref{peta}) into Eq. (\ref{tsaent}) directly leads
to the formula
\begin{equation}
(1-q)H_{q}(A_{\eta})=\eta^{q}\left(\sum\nolimits_{a}p(a)^{q}-1\right)+\eta^{q}+(1-\eta)^{q}-1
\ . \label{rlem01}
\end{equation}
Dividing Eq. (\ref{rlem01}) by $(1-q)$ gives the claim
(\ref{rlem0}). Similarly to Eq. (\ref{rlem01}), we further
write the term $(1-q)H_{q}(A_{\eta\eta})$ as
\begin{equation}
\eta^{2q}{\left(\sum_{a}p(a)^{q}-1\right)}+\eta^{q}(1-\eta)^{q}{\left(\sum_{b}p(b)^{q}+\sum_{c}p(c)^{q}-2\right)}
+\eta^{2q}+2\eta^{q}(1-\eta)^{q}+(1-\eta)^{2q}-1
{\,}. \label{rlem001}
\end{equation}
By the identity
$\eta^{2q}+2{\eta}^{q}(1-\eta)^{q}+(1-\eta)^{2q}-1=\bigl(\eta^{q}+(1-\eta)^{q}+1\bigr)\bigl(\eta^{q}+(1-\eta)^{q}-1\bigr)$,
we get the claim (\ref{rlem00}) from Eq. (\ref{rlem001}) after
dividing by $(1-q)$. The standard case is recovered in the limit
$q\to1$. $\blacksquare$

The second summand in the right-hand side of Eq. (\ref{rlem0}) can
easily be checked with any deterministic probability distribution.
If the initial distribution $\{p(a)\}$ is deterministic, then the
deformed distribution (\ref{peta}) includes only two nonzero
probabilities, namely $\eta$ and $(1-\eta)$. As $H_{q}(A)=0$, the
right-hand side of Eq. (\ref{rlem0}) actually gives the binary
$q$-entropy. Similarly, the third summand in the right-hand side
of Eq. (\ref{rlem00}) could be checked with three deterministic
probability distributions.

Originally, the Braunstein--Caves inequality was formulated with
use of the conditional entropy and its generic properties
\cite{BC88}. The writers of Ref. \cite{cerf97} derived entropic
Bell inequalities by considering the so-called entropy Venn
diagrams. The entropy of $A$ conditional on knowing $B$ is defined
as \cite{CT91}
\begin{equation}
H_{1}(A|B):=\sum\nolimits_{b} p(b){\,}H_{1}(A|b)
=-\sum\nolimits_{a}\sum\nolimits_{b} p(a,b){\,}\ln{p}(a|b)
\ , \label{cshen}
\end{equation}
where $H_{1}(A|b):=-\sum_{a}p(a|b)\ln{p}(a|b)$ and
$p(a|b)=p(a,b){\,}p(b)^{-1}$ according to the Bayes rule. The
quantity (\ref{cshen}) will be referred to as the standard
conditional entropy. Further, we will use its $q$-entropic
extension. By means of the particular functional
\begin{equation}
H_{q}(A|b):=-\sum_{a}p(a|b)^{q}\ln_{q}p(a|b)
=\sum_{a}p(a|b){\,}\ln_{q}\frac{1}{p(a|b)}
\ , \label{pcen}
\end{equation}
one defines the conditional $q$-entropy \cite{sf06,rastkyb}
\begin{equation}
H_{q}(A|B):=\sum\nolimits_{b} p(b)^{q}{\,}H_{q}(A|b)
\ . \label{qshen}
\end{equation}
In the limit $q\to1$, this definition is reduced to Eq.
(\ref{cshen}). The above entropic measures with $q=2$ have been
used in Ref. \cite{vajda68} for estimating the error probability
on checking statistical hypotheses. Below, we will extensively use
the following properties of the entropic function (\ref{qshen}).
For all $q>0$, the entropy (\ref{qshen}) satisfies
\begin{equation}
H_{q}(A,B)=H_{q}(B|A)+H_{q}(A)
=H_{q}(A|B)+H_{q}(B)
\ . \label{chrl}
\end{equation}
This formula expresses the chain rule for the conditional
$q$-entropy \cite{sf06}. It can easily be derived in line with the
definitions (\ref{tsaln}) and (\ref{qshen}) by means of the
identity
\begin{equation}
\ln_{q}(xy)=\ln_{q}x+x^{1-q}\ln_{q}y
\ . \label{xyid}
\end{equation}
The mutual information is widely used in
information theory \cite{CT91}. Similarly to
the standard case, the mutual $q$-information can be defined as
\cite{sf06}
\begin{equation}
I_{q}(A:B):=H_{q}(A)-H_{q}(A|B)
\ . \label{minq}
\end{equation}
For $q=1$, we have the standard mutual information
$I_{1}(A:B)=H_{1}(A)-H_{1}(A|B)$. Using normalized Tsallis
entropies, the corresponding mutual information was introduced in
Ref. \cite{tyam01}. We can rewrite (\ref{minq}) in the form
\begin{equation}
I_{q}(A:B)=I_{q}(B:A)=H_{q}(A)+H_{q}(B)-H_{q}(A,B)
\ ,  \label{minq1}
\end{equation}
since $H_{q}(A|B)=H_{q}(A,B)-H_{q}(B)$ by Eq. (\ref{chrl}). So, the
quantity (\ref{minq}) is symmetric in its entries. Quantum
violations of the Clauser--Horne--Shimony--Holt inequality is
limited from above by the Tsirel'son bound \cite{ts80}. This bound
can be derived from the assumption that the chain rule holds for a
generalized mutual information proposed in Ref. \cite{wm12}.

The chain rule (\ref{chrl}) can further be extended to more than
two variables. According to theorem 2.4 of Ref. \cite{sf06}, one
obeys
\begin{equation}
H_{q}(A_{1},A_{2},\ldots,A_{n})=\sum\nolimits_{j=1}^{n}H_{q}(A_{j}|A_{j-1},\ldots,A_{1})
\ . \label{ehrl}
\end{equation}
Using Eq. (\ref{chrl}) and non-negativity of the conditional
$q$-entropy, we immediately obtain
\begin{equation}
H_{q}(A)\leq{H}_{q}(A,B)
\ , \qquad
H_{q}(B)\leq{H}_{q}(A,B)
\ . \label{habq}
\end{equation}
In the next section, we will also use inequalities of the
following form.

\newtheorem{t21}[t20]{Lemma}
\begin{t21}\label{lem1}
For real $q\geq1$ and integer $n\geq1$, the conditional
$q$-entropy satisfies
\begin{equation}
H_{q}(A|B_{1},\ldots,B_{n-1},B_{n})
\leq{H}_{q}(A|B_{1},\ldots,B_{n-1})
\ . \label{rlem1}
\end{equation}
\end{t21}

{\bf Proof.} Let us assume $q>1$. First, we prove the claim for
$n=2$. The conditional $q$-entropy $H_{q}(A|B,C)$ can be rewritten
as
\begin{equation}
H_{q}(A|B,C)=\sum_{ab} p(b)^{q}
\sum_{c} \left(\frac{p(b,c)}{p(b)}\right)^{q}
f_{q}\bigl(p(a|b,c)\bigr)
\ , \label{habc}
\end{equation}
where the function $f_{q}(x):=\bigl(x^{q}-x\bigr)/(1-q)$ is
concave. Since $\sum_{c}p(c|b)=1$, we have
$p(c|b)^{q}\leq{p}(c|b)$ for $q\geq1$. So, the sum with respect to
$c$ obeys
\begin{equation}
\sum\nolimits_{c} p(c|b)^{q} f_{q}\bigl(p(a|b,c)\bigr)
\leq
\sum\nolimits_{c} p(c|b){\,}f_{q}\bigl(p(a|b,c)\bigr)
\leq
f_{q}\left(\sum\nolimits_{c} p(c|b){\,}p(a|b,c)\right)
{\,}, \label{sumc}
\end{equation}
due to Jensen's inequality. As the numbers
$p(c|b){\,}p(a|b,c)=p(b,c){\,}p(b)^{-1}p(a,b,c){\,}p(b,c)^{-1}=p(a,b,c){\,}p(b)^{-1}$
are summarized to $p(a,b){\,}p(b)^{-1}=p(a|b)$, the right-hand
side of Eq. (\ref{sumc}) reads $f_{q}\bigl(p(a|b)\bigr)$.
Combining this with Eq. (\ref{habc}) then gives
\begin{equation}
H_{q}(A|B,C)\leq\sum\nolimits_{ab} p(b)^{q}f_{q}\bigl(p(a|b)\bigr)=H_{q}(A|B)
\ . \label{sumb}
\end{equation}
By a parallel argument, we easily have the case $n=1$, namely
\begin{equation}
H_{q}(A|B)\leq{H}_{q}(A)
\ . \label{suma}
\end{equation}
The proof of Eq. (\ref{rlem1}) is completed by an extension with
respect to $n$. The case $q=1$ can be recovered by repeating the
above reasons with the concave function $f_{1}(x)=-x\ln{x}$.
$\blacksquare$

Note that the formula (\ref{suma}) implies positivity of the
mutual $q$-information (\ref{minq}) for all $q\geq1$. There exists
another form of the conditional $q$-entropy \cite{sf06}. However,
this form does not succeed some useful relations including the
chain rule. Properties of both forms of the conditional
$q$-entropy are discussed in the papers \cite{sf06,rastkyb}. The
Fano inequality in terms of $q$-entropies and some of its
applications are considered in Refs. \cite{sf06,rastmpag}. We will
use the conditional $q$-entropy of order $q\geq1$ for expressing
inequalities of Bell's type.

\section{Marginal scenarios and Bell inequalities}\label{sec3}

The notion of marginal scenarios provides a general way
to treat the noncontextuality of probability distributions
\cite{rchtf12,chfr13}. In a marginal problem, we ask whether a
given family of marginal distributions for some set of random
variables arises from some joint distribution of these variables
\cite{chfr13}. Both Bell scenarios and contextuality scenarios can
be unified in the following way \cite{rchtf12}. Let
$\bigl\{X_{1},\ldots,X_{n}\bigr\}$ be a finite set of observables,
and let $\mc=\bigl\{S_{1},\ldots,S_{|\mc|}\bigr\}$ be a family of
subsets $S_{i}\subseteq\bigl\{X_{1},\ldots,X_{n}\bigr\}$. Such
subsets are assumed to be comprised from commuting observables. In
other words, each subset contains jointly measurable quantities.
Hence, the two conditions $S\in\mc$ and $S^{\prime}\subseteq{S}$
must imply $S^{\prime}\in\mc$. When the family $\mc$ obeys this
implication, we call it ``marginal scenario''. For a formal
consistency, the empty set $\emptyset$ is assumed to be included
into $\mc$.

From the physical viewpoint, one obtains some joint measurement
statistics for each $S\in\mc$. In real experiments, physicists
usually deal with a collection of pairs of compatible observables.
Suppose that $\{X,Y\}\in\mc$. By $\pbr(x,y|X,Y)$, we denote the
probability of obtaining the outcomes $x$ for $X$ and $y$ for $Y$
in their joint measurement. A similar notation will be used for
more than two compatible observables. Note that the notation
$\pbr(x,y|X,Y)$ assumes the specific physical context. In this
sense, such probabilities should be distinguished from usual
conditional probabilities. The introduced probabilities are used
to pose formally criteria that given probabilistic model is not
contextual \cite{chfr13}. An approach based on the algebraic
language has been developed by Abramsky and Brandenburger
\cite{abrams11}.

Within an intuitive approach, we assign some hidden variable
$\lambda$ to any physical model. It is assumed that this variable
completely predetermines the future behavior. If the actual value
of $\lambda$ was known, the probabilities $p_{X}(x|\lambda)$ of
each observable $X$ are assumed to be independent of measurement
statistics of all other observables \cite{rchtf12}. Hence, for
mutually compatible $X$ and $Y$ we can write
\begin{equation}
\pbr(x,y|X,Y)=
\sum\nolimits_{\lambda}\varrho(\lambda){\,}p_{X}(x|\lambda){\,}p_{Y}(y|\lambda)
\ . \label{xylm}
\end{equation}
Here, unknown quantities $\varrho(\lambda)$ must obey
$\varrho(\lambda)\geq0$ and $\sum_{\lambda}\varrho(\lambda)=1$.
Similarly to Eq. (\ref{xylm}), we can deal with more than two
compatible observables. The noncontextuality of a given model in
marginal scenario $\mc$ implies the existence of a joint
probability distribution
\begin{equation}
\pbr(x_{1},\ldots,x_{n}|X_{1},\ldots,X_{n})=p(x_{1},\ldots,x_{n})
\ , \label{xxjp}
\end{equation}
which marginalizes to the model distributions for all $S\in\mc$
\cite{chfr13,abrams11}. We then aim to decide, whether the
considered probabilistic model obeys this criterion. It can be
rewritten in terms of mean values or entropic functions.

Original Bell inequalities \cite{bell64} were written in terms of
mean values. Results of such a kind usually pertain to
experiments, which probe entanglement between spacelike separated
subsystems. The CHSH scenario is probably the most known setup.
Let observables $A$ and $A^{\prime}$ be used for one subsystem,
and let observables $B$ and $B^{\prime}$ be used for other. Both
the pairs $\{A,A^{\prime}\}$ and $\{B,B^{\prime}\}$ are not
jointly measurable. On the other hand, each element of
$\{A,A^{\prime}\}$ is compatible with each element of
$\{B,B^{\prime}\}$, since they are related to different
subsystems. So, the marginal scenario includes the four singletons
$\{A\}$, $\{A^{\prime}\}$, $\{B\}$, $\{B^{\prime}\}$, and the four
pairs $\{A,B\}$, $\{A,B^{\prime}\}$, $\{A^{\prime},B\}$,
$\{A^{\prime},B^{\prime}\}$. In the usual CHSH scenario, each of
the observables has two possible outcomes. Let outcomes be
rescaled to $\pm1$. The existence of a joint probability
distribution for this scenario then leads to the CHSH inequality
\cite{chsh69}
\begin{equation}
\langle{A}B^{\prime}\rangle+\langle{A}^{\prime}B^{\prime}\rangle+\langle{A}^{\prime}B\rangle
-\langle{A}B\rangle\leq2
\ . \label{chshin}
\end{equation}
Quantum mechanics predicts that the left-hand side of Eq.
(\ref{chshin}) can increase up to $2\sqrt{2}$ \cite{ts80}.
Violations of Eq. (\ref{chshin}) have been tested in experiments
\cite{agr82,adr82}. Similarly, we formulate the scenario with
arbitrary number of outcomes for observables. Assuming the
existence of a joint probability distribution, Braunstein and
Caves derived entropic inequality \cite{BC88}
\begin{equation}
H_{1}(A|B)\leq{H}_{1}(A|B^{\prime})+H_{1}(B^{\prime}|A^{\prime})
+H_{1}(A^{\prime}|B)
\ . \label{comb31}
\end{equation}
The conditional entropy is asymmetric in its entries. The authors
of Ref. \cite{rchtf12} rewrite Eq. (\ref{comb31}) in terms of the
symmetrical mutual information, namely
\begin{equation}
I_{1}(A:B^{\prime})+I_{1}(A^{\prime}:B^{\prime})+I_{1}(A^{\prime}:B)-I_{1}(A:B)
\leq{H}_{1}(A^{\prime})+H_{1}(B^{\prime})
\ . \label{chshmi}
\end{equation}
In a structure, the information-theoretic inequality
(\ref{chshmi}) is similar to the usual CHSH inequality
(\ref{chshin}). When we apply Eq. (\ref{chshmi}) to test the
nonlocality of a probability distribution, the following
symmetries should be taken into account. By a permutation, the
right-hand side of this inequality can be rewritten with every
pair of compatible observables.

Unlike the CHSH scenario, the KCBS scenario \cite{kly08} is not
associated with correlations between the measurements on different
subsystems. The latter pertain to the measurements statistics for
a single system. Here, we deal with five quantities $X_{1}$,
$X_{2}$, $X_{3}$, $X_{4}$, $X_{5}$, such that each pair
$\{X_{j},X_{j+1}\}$ is jointly measurable. If quantities take
values $\pm1$, then the existence of a joint probability
distribution leads to the pentagram inequality \cite{kly08}
\begin{equation}
\sum\nolimits_{j=1}^{5} \langle{X}_{j}X_{j+1}\rangle\geq-3
\ . \label{penin}
\end{equation}
The corresponding entropic formulation is expressed as \cite{krk12}
\begin{equation}
H_{1}(X_{1}|X_{5})\leq{H}_{1}(X_{1}|X_{2})+H_{1}(X_{2}|X_{3})+H_{1}(X_{3}|X_{4})+H_{1}(X_{4}|X_{5})
\ . \label{ca151}
\end{equation}
The writers of Ref. \cite{rchtf12} gave this inequality in other
form known as the entropic Klyachko inequality. Advantages of
entropic formulations are the following. First, they can handle
any finite number of outcomes. Second, the entropic approach
allows to study more realistic cases with detection inefficiencies
\cite{rchtf12}. Further, we will consider the following two
models.

In the first model, two compatible observables are measured
jointly by a single detector. By $\eta\in[0;1]$, we quantify a
detection efficiency. The no-click event is represented by
additional outcome $(\varnothing,\varnothing)$. The new
probability distribution includes the probabilities \cite{rchtf12}
\begin{align}
\pbr^{(\eta)}(x_{j},x_{j+1}|X_{j},X_{j+1})&=\eta{\,}\pbr(x_{j},x_{j+1}|X_{j},X_{j+1})
\ , \label{pretx}\\
\pbr^{(\eta)}(\varnothing,\varnothing|X_{j},X_{j+1})&=1-\eta
\ , \label{pret0}
\end{align}
where $x_{j},x_{j+1}\in\{-1,+1\}$. This probability distribution
marginalizes to the single-observable distribution
\begin{equation}
\pbr^{(\eta)}(x_{j}|X_{j})=\eta{\,}\pbr(x_{j}|X_{j})
\ , \qquad  \pbr^{(\eta)}(\varnothing|X_{j})=1-\eta
\ . \label{pret0m}
\end{equation}
In this model, the no-click event occurs for both observables
simultaneously with the probability (\ref{pret0}). As shown in
Ref. \cite{rchtf12}, the entropic Klyachko inequality merely
scales by $\eta$. Thus, the inequality has a violation for all
$\eta>0$. Violations take place in the same cases, for which the
inequality with $\eta=1$ is violated. We will further show that
these properties remain valid for the corresponding $q$-entropic
inequalities.

In the second model, the joint measurement of $X_{j}$ and
$X_{j+1}$ is performed by two detectors. We assume that each of
detectors has an efficiency of $\eta\in[0;1]$. It can be realized
within some sequential scheme with a nondemolition measurement in
the first detector \cite{rchtf12}. For any jointly measurable
pair, one writes the probabilities
\begin{align}
\pbr^{(\eta\eta)}(x_{j},x_{j+1}|X_{j},X_{j+1})&=\eta^{2}{\,}\pbr(x_{j},x_{j+1}|X_{j},X_{j+1})
\ , \label{pretxx}\\
\pbr^{(\eta\eta)}(x_{j},\varnothing|X_{j},X_{j+1})&=\eta(1-\eta){\,}\pbr(x_{j}|X_{j})
\ , \label{pretx0}\\
\pbr^{(\eta\eta)}(\varnothing,\varnothing|X_{j},X_{j+1})&=(1-\eta)^{2}
\ , \label{pret00}
\end{align}
where $\pbr^{(\eta\eta)}(\varnothing,x_{j+1}|X_{j},X_{j+1})$ is
expressed similarly to Eq. (\ref{pretx0}). This probability
distribution also marginalizes to the single-observable
distribution (\ref{pret0m}). In this model, the required detection
efficiency for witnessing quantum violations turned out be very
high, $\eta\approx0.995$ \cite{rchtf12}. In the following, we will
consider this issue for $q$-entropic inequalities of the Bell
type.

The CHSH and KCBS scenarios are both particular cases of the
$n$-cycle \cite{lsw11,aqbcc12}. This notion is defined
for any number $n\geq3$ of observables $X_{1},\ldots,X_{n}$ in a
cyclic configuration. We demand that two observables $X_{j}$
and $X_{j+1}$ be jointly measurable for all $j=1,\ldots,n$. The
complete characterization of the $n$-cycle scenario has been given
for dichotomic observables, when possible outcomes are $\pm1$. Let
each of $n$ factors $\gamma_{j}$ be either $-1$ or $+1$, and let
the total number of $\gamma_{j}=-1$ be odd. Then the
noncontextuality implies \cite{aqbcc12}
\begin{equation}
\sum\nolimits_{j=1}^{n} \gamma_{j}\langle{X}_{j}X_{j+1}\rangle\leq{n}-2
\ . \label{cncyl}
\end{equation}
All $2^{n-1}$ inequalities of the form (\ref{cncyl}) characterize
the $n$-cycle noncontextual polytope \cite{aqbcc12}. The CHSH
inequality (\ref{chshin}) is an example of Eq. (\ref{cncyl}) for
$n=4$. Entropic formulations for the $n$-cycle scenario are
examined in Refs. \cite{rchtf12,chfr13}.

\section{Entropic inequalities for the CHSH scenario}\label{sec4}
\noindent In this section, we formulate Bell's theorem in terms of
the conditional $q$-entropies for the CHSH scenario. The
$q$-entropic inequalities will be derived from the existence of
joint probability distribution
$p\bigl(a,b^{\prime},a^{\prime},b\bigr)$. This joint distribution
should marginalize to the model distributions for all jointly
measurable pairs. For instance, for the pair $\{A,B\}$ we have
\begin{equation}
p(a,b)=\sum\nolimits_{b^{\prime}a^{\prime}} p(a,b^{\prime},a^{\prime},b)
\ , \label{pjpb}
\end{equation}
and similarly for other jointly measurable subsets. Due to
relations of the form (\ref{habq}), we write
\begin{equation}
H_{q}(A,B)\leq{H}_{q}(A,B^{\prime},A^{\prime},B)=
H_{q}(A|B^{\prime},A^{\prime},B)+H_{q}(B^{\prime}|A^{\prime},B)
+H_{q}(A^{\prime}|B)+H_{q}(B)
\ . \label{comb0}
\end{equation}
Here, the entropy $H_{q}(A,B^{\prime},A^{\prime},B)$ was expressed
with respect to the chain rule (\ref{ehrl}). Subtracting
$H_{q}(B)$ and using Eq. (\ref{chrl}), one further obtains
\begin{equation}
H_{q}(A|B)\leq{H}_{q}(A|B^{\prime},A^{\prime},B)+H_{q}(B^{\prime}|A^{\prime},B)
+H_{q}(A^{\prime}|B)
\ . \label{comb1}
\end{equation}
According to Lemma \ref{lem1}, for $q\geq1$ we write
\begin{equation}
H_{q}(A|B^{\prime},A^{\prime},B)\leq{H}_{q}(A|B^{\prime})
\ , \qquad
H_{q}(B^{\prime}|A^{\prime},B)\leq{H}_{q}(B^{\prime}|A^{\prime})
\ . \label{comb2}
\end{equation}
Combining these relations with Eq. (\ref{comb1}), we have arrived
at the entropic inequality
\begin{equation}
H_{q}(A|B)\leq{H}_{q}(A|B^{\prime})+H_{q}(B^{\prime}|A^{\prime})
+H_{q}(A^{\prime}|B)
\ , \label{comb3}
\end{equation}
which holds for $q\geq1$. Predictions of quantum mechanics
sometimes lead to a violation of Eq. (\ref{comb3}). For $q=1$,
this formula is reduced to the Braunstein--Caves inequality
(\ref{comb31}). Using the conditional $q$-entropies, we herewith
obtained a one-parametric extension of the main result of Ref.
\cite{BC88}. In terms of the mutual $q$-information, for $q\geq1$
we also have
\begin{equation}
I_{q}(A:B^{\prime})+I_{q}(A^{\prime}:B^{\prime})+I_{q}(A^{\prime}:B)-I_{q}(A:B)
\leq{H}_{q}(A^{\prime})+H_{q}(B^{\prime})
\ . \label{chshmq}
\end{equation}
It follows from Eq. (\ref{comb3}) by immediate use of the
definition (\ref{minq}). Similarly to Eq. (\ref{chshmi}), we
should keep in mind possible permutations of the jointly
measurable pairs in Eq. (\ref{chshmq}).

\begin{figure}
\includegraphics[width=8.5cm]{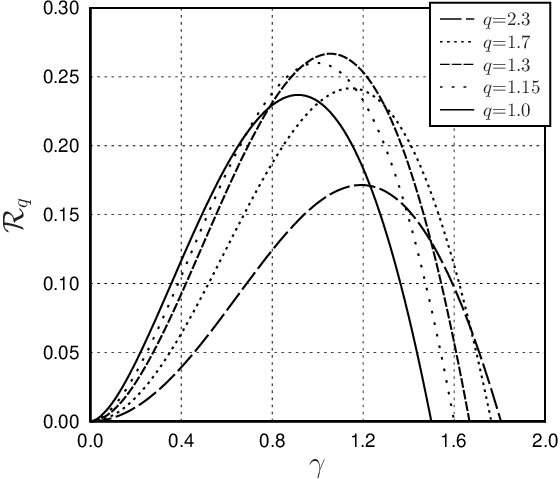}
\caption{\label{fig1}The relative quantity $\tca_{q}$ versus
$\gamma$ in the case of Eq. (\ref{rwcq}) for
$q=1.0;1.15;1.3;1.7;2.3$. For each $q$, only positive values of
$\tca_{q}$ are shown.}
\end{figure}

To observe violations of Eq. (\ref{comb3}), we will deal with the
four observables
\begin{align}
\msa &=\va\cdot\vec{\bsg}\otimes\pen \ , &  \msa^{\prime} &=\va^{\prime}\cdot\vec{\bsg}\otimes\pen \ , &
\label{aap}\\
\msb &=\pen\otimes\vb\cdot\vec{\bsg} \ , &  \msb^{\prime} &=\pen\otimes\vb^{\prime}\cdot\vec{\bsg} \ . &
\label{bbp}
\end{align}
Here, the three-dimensional vectors $\va$, $\va^{\prime}$, $\vb$,
and $\vb^{\prime}$ are unit; the $\vec{\bsg}$ is the vector of
Pauli matrices. Violations of Eq. (\ref{comb3}) can be
characterized by the quantity
\begin{equation}
\ca_{q}=H_{q}(A|B)-H_{q}(A|B^{\prime})-H_{q}(B^{\prime}|A^{\prime})
-H_{q}(A^{\prime}|B)
\ . \label{caAB}
\end{equation}
Following Ref. \cite{BC88}, we consider coplanar three-dimensional
vectors $\va$, $\vb^{\prime}$, $\va^{\prime}$, and $\vb$, with the
angles
$\measuredangle(\va,\vb^{\prime})=\measuredangle(\vb^{\prime},\va^{\prime})=\measuredangle(\va^{\prime},\vb)=\gamma/3$
and $\measuredangle(\va,\vb)=\gamma$. The initial state of two
spin-$1/2$ systems is the state of zero total spin, namely
\begin{equation}
|\Phi\rangle=\frac{1}{\sqrt{2}}{\>}\Bigl(|0\rangle\otimes|1\rangle-|1\rangle\otimes|0\rangle\Bigr)
\ . \label{szts}
\end{equation}
In Eq. (\ref{szts}), the quantization axis is completely
arbitrary. With such a choice, the characteristic quantity
(\ref{caAB}) can be rewritten as
\begin{equation}
\ca_{q}=H_{q}(A|B)-3H_{q}(B^{\prime}|A^{\prime})
\ . \label{rwcq}
\end{equation}
Here, the first term corresponds to the angle $\gamma$ between two
unit vectors, and the second one corresponds to the angle
$\gamma/3$ between two unit vectors. Positive values of $\ca_{q}$
imply violations of the locality hypothesis. It is useful to
measure these positive values with a natural scale of entropic
values. So, we will relate $\ca_{q}$ with the number $\ln_{q}2$,
which represents the maximal binary $q$-entropy. That is, the
results are reported in terms of the relative quantity
\begin{equation}
\tca_{q}:=\left(\ln_{q}2\right)^{-1}\ca_{q}
\ . \label{relq}
\end{equation}
Figure \ref{fig1} presents violations of Eq. (\ref{comb3}) for
various $q\geq1$. For comparison, we include the standard case
$q=1$, when the maximum is equal to $0.2369$ \cite{BC88} and
reached for $\gamma=0.9141$. With increase of $q$, the curve
maximum goes to larger values of $\gamma$. One shows some
extension of the domain, for which $\tca_{q}>0$. The inequality
(\ref{comb3}) is actually violated for one values of $q$ and is
not violated for other, including $q=1$. Here, we can recall
symmetries of Eqs. (\ref{chshmi}) and (\ref{chshmq}) with respect
to permutations of the four measurable pairs. In the considered
example, however, such permutations do not give new detectable
cases for fixed $q$. Due to invariance of the state (\ref{szts}),
we have relations of the form $H_{q}(A|B)=H_{q}(B|A)$, which
depend only on the angle between two unit vectors. Thus, the
$q$-entropic inequalities can detect the nonlocality of some
probability distributions that cannot be detected by Eq.
(\ref{comb31}).

In general, entropic inequalities give only necessary criteria for
the locality or noncontextualuity. In this sense, the $q$-entropic
inequalities provide more powerful criteria. In the dichotomic
CHSH scenario, the author of Ref. \cite{rch13} has recently shown
the following. Adding a shared randomness in the experimental
setup, the Braunstein--Caves inequalities turn to be sufficient.
Extending the depolarization protocol of Ref. \cite{mag06}, the
sufficiency can be stated for any $n$-cycle with dichotomic
outcomes. There exists also an argument without the depolarization
procedure \cite{rch13}. Using the entire family of $q$-entropic
inequalities provides a complementary way, which can be essential
with more than two outcomes.

It is easy to write the $q$-entropic inequalities for the
$n$-cycle scenario. Here, each pair $\{X_{j},X_{j+1}\}$ is jointly
measurable. We suppose that there exist a joint probability
distribution $p(x_{1},x_{2},\ldots,x_{n})$, which marginalizes to
two-observable distributions of the form
\begin{equation}
p(x_{1},x_{2})
=\sum_{x_{j}:{\>}j\neq1,2} p(x_{1},x_{2},\ldots,x_{n})
\ . \label{akaj}
\end{equation}
Assuming this, we extend Eq. (\ref{comb3}) in the following way.
For $q\geq1$, one has
\begin{equation}
H_{q}(X_{1}|X_{n})\leq
\sum\nolimits_{j=1}^{n-1}
H_{q}(X_{j}|X_{j+1})
\ . \label{rthm}
\end{equation}
This formula can be derived by means of obvious extension of the
reasons from Eqs. (\ref{comb0})--(\ref{comb3}). We refrain from
presenting the details here. Using Eqs. (\ref{chrl}) and
(\ref{minq}), we could rewrite the inequality (\ref{rthm}) with
use of the joint $q$-entropies or the mutual $q$-informations. The
former is essential in studying models of detection
inefficiencies. We consider this issue in the next section. To
compare Eq. (\ref{rthm}) with predictions of quantum theory, we
will use an immediate extension of Eq. (\ref{caAB}). For $q=1$,
such a quantity for the KCBS scenario was considered in Ref.
\cite{krk12}. The inequality (\ref{rthm}) is then rewritten as
$\ca_{q}\leq0$. If predictions of quantum mechanics do sometimes
lead to strictly positive $\ca_{q}$, then the noncontextuality
hypothesis fails. In such a case, the quantity $\ca_{q}$
characterizes an amount of violation of the inequality
(\ref{rthm}). As was argued in Ref. \cite{krk12}, violation of the
inequality (\ref{rthm}) implies violation of the corresponding
pentagram inequality of Ref. \cite{kly08}, but the converse is not
true. Such findings could be verified in appropriate experiments.

\section{Entropic inequalities for the KCBS scenario}\label{sec5}

In this section, we examine $q$-entropic inequalities of the Bell
type for the KCBS scenario. In the case $n=5$ and $q\geq1$, the entropic
inequality (\ref{rthm}) reads
\begin{equation}
H_{q}(X_{1}|X_{5})\leq{H}_{q}(X_{1}|X_{2})+H_{q}(X_{2}|X_{3})+H_{q}(X_{3}|X_{4})+H_{q}(X_{4}|X_{5})
\ . \label{comb5}
\end{equation}
We also recall symmetries of such inequalities with respect to
acceptable permutations. Following Refs. \cite{rchtf12,krk12}, we
consider projectors of the form $|X_{k}\rangle\langle{X}_{k}|$
with the eigenvectors
\begin{align}
 & |X_{1}\rangle=\frac{1}{\sqrt{2}\cos\alpha}
\begin{pmatrix}
\sqrt{\cos2\alpha} \\
\sin\alpha \\
\cos\alpha
\end{pmatrix}
{\,}, \qquad
|X_{2}\rangle=
\begin{pmatrix}
0 \\
\cos\alpha \\
-\sin\alpha
\end{pmatrix}
{\,}, \qquad
|X_{3}\rangle=
\begin{pmatrix}
1 \\
0 \\
0
\end{pmatrix}
{\,}, \label{123a}\\
 & |X_{4}\rangle=
\begin{pmatrix}
0 \\
\cos\alpha \\
\sin\alpha
\end{pmatrix}
{\,}, \qquad
|X_{5}\rangle=\frac{1}{\sqrt{2}\cos\alpha}
\begin{pmatrix}
\sqrt{\cos2\alpha} \\
\sin\alpha \\
-\cos\alpha
\end{pmatrix}
{\,}, \label{45a}
\end{align}
where $\alpha\in(0;\pi/4)$. The five vectors satisfy orthogonality
conditions
\begin{equation}
\langle{X}_{1}|X_{2}\rangle=
\langle{X}_{2}|X_{3}\rangle=
\langle{X}_{3}|X_{4}\rangle=
\langle{X}_{4}|X_{5}\rangle=
\langle{X}_{5}|X_{1}\rangle=0
\ . \label{ortc}
\end{equation}
The two projectors $|X_{k}\rangle\langle{X}_{k}|$ and
$|X_{k+1}\rangle\langle{X}_{k+1}|$ are jointly measurable for all
$k=1,2,3,4,5$. Eigenvalues $1$ and $0$ of the projector
$|X_{k}\rangle\langle{X}_{k}|$ respectively correspond to outcomes
``yes'' and ``no'', when measured quantum state passes the test of
being the state $|X_{k}\rangle$. The vectors
(\ref{123a})--(\ref{45a}) also obey
$\langle{X}_{1}|X_{4}\rangle=\langle{X}_{5}|X_{2}\rangle$ and
$\langle{X}_{1}|X_{3}\rangle=\langle{X}_{5}|X_{3}\rangle$.
Further, we write the pre-measurement state as
\begin{equation}
|\psi\rangle=
\begin{pmatrix}
\sin\theta \\
\cos\theta \\
0
\end{pmatrix}
{\>}, \label{meps}
\end{equation}
for which $\langle{X}_{1}|\psi\rangle=\langle{X}_{5}|\psi\rangle$
and $\langle{X}_{2}|\psi\rangle=\langle{X}_{4}|\psi\rangle$. Some
intuitive reasons for such a configuration are briefly discussed
in Ref. \cite{krk12}.

With the pre-measurement state $|\psi\rangle$, the
observation of $X_{k}$ leads to the outcomes $x_{k}=1$ and
$x_{k}=0$ with probabilities $|\langle{X}_{k}|\psi\rangle|^{2}$
and $1-|\langle{X}_{k}|\psi\rangle|^{2}$, respectively. According
to the projection postulate, the normalized post-measurement state
is $|X_{k}\rangle$ for $x_{k}=1$ and
\begin{equation}
\left(1-|\langle{X}_{k}|\psi\rangle|^{2}\right)^{-1/2}\Bigl\{|\psi\rangle-|X_{k}\rangle\langle{X}_{k}|\psi\rangle\Bigr\}
\label{post0}
\end{equation}
for $x_{k}=0$. Hence, the context for next observations is
determined. If the next observation is $X_{j}$, we calculate the
conditional probabilities and, further, the corresponding entropy
$H_{q}(X_{j}|X_{k})$. In this quantum-mechanical way, one
evaluates the characteristic quantity
\begin{equation}
\ca_{q}=H_{q}(X_{1}|X_{5})-H_{q}(X_{1}|X_{2})-H_{q}(X_{2}|X_{3})-H_{q}(X_{3}|X_{4})-H_{q}(X_{4}|X_{5})
\ . \label{ca15}
\end{equation}

\begin{table}
\begin{center}
\caption{\label{tab1}The maximal values of $\ca_{q}$ and
$\tca_{q}$ for several $q$.} \vskip0.1cm
\begin{tabular}{|c|c|c|c|c|c|c|c|c|c|c|c|c|}
\hline
$q$ & 1.0 & 1.1 & 1.2 & 1.4 & 1.6 & 1.8 & 2.0 & 2.5 & 3.0 & 5.0 & 8.0 & 11.0 \\
\hline
$\max\ca_{q}$ & 0.0631 & 0.0779 & 0.0898 & 0.1049 & 0.1111 & 0.1113 & 0.1079 & 0.0924 & 0.0759 &  0.0383 & 0.0212 & 0.0146 \\
$\max\tca_{q}$ & 0.0911 & 0.1164 & 0.1387 & 0.1733 & 0.1960 & 0.2093 & 0.2157 & 0.2143 & 0.2024 & 0.1632 & 0.1494 & 0.1462 \\
$\alpha_{\rm{max}}$ & 0.1698 & 0.1802 & 0.1880 & 0.1987 & 0.2051 & 0.2085 & 0.2099 & 0.2067 & 0.1982 & 0.1557 & 0.1205 & 0.1017 \\
$\theta_{\rm{max}}$ & 0.2366 & 0.2684 & 0.2943 & 0.3327 & 0.3585 & 0.3761 & 0.3880 & 0.4014 & 0.3996 & 0.3345 & 0.2639 & 0.2247 \\
\hline
\end{tabular}
\end{center}
\end{table}

\begin{figure}
\includegraphics[width=8.5cm]{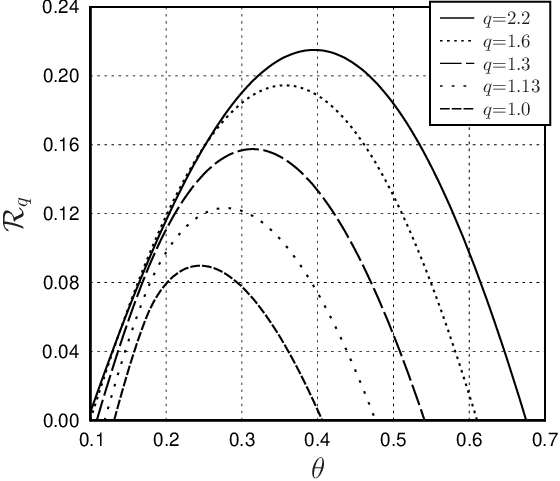}
\caption{\label{fig2}The relative quantity $\tca_{q}$ versus
$\theta$ in the case of Eq. (\ref{ca15}) for $\alpha=0.1885$ and
$q=1.0;1.13;1.3;1.6;2.2$. For each $q$, only positive values of
$\tca_{q}$ are shown.}
\end{figure}

The inequality (\ref{comb5}) implies $\ca_{q}\leq0$. The main
result is its violations for certain values of the parameters
$\alpha$ and $\theta$. We do not solve analytically the problem of
finding a joint parametric domain, in which $\ca_{q}>0$. For given
parameters, however, the quantity $\ca_{q}$ is easy to numerical
estimation. Some numerical results are summarized below. Here, we
will again use the quantity rescaled according to Eq.
(\ref{relq}). In Table \ref{tab1}, the maximal values of $\ca_{q}$
and $\tca_{q}$ are shown for several values of the parameter $q$.
The values $\alpha_{\rm{max}}$ and $\theta_{\rm{\max}}$, which
correspond to the maximal violation, are given as well. In
relative entropic size, the maximal violation of Eq. (\ref{comb5})
is sufficiently large for all the presented values of $q$. The
standard case $q=1$ was previously reported in Ref. \cite{krk12}.
For convenience of comparing with values $q>1$, we insert this
case in the table. As we see in Table \ref{tab1}, the values
$\alpha_{\rm{max}}$ and $\theta_{\rm{\max}}$ depend on $q$. In
given experimental setting, some fixed value of $\alpha$ and few
values of $\theta$ would be rather available. On Fig. \ref{fig2},
a dependence of $\tca_{q}$ on $\theta$ is given for
$\alpha=0.1885$ and five values of the parameter $q$. We see that
violation of Eq. (\ref{comb5}) is significant for many values
$q\geq1$. Curves of Fig. \ref{fig2} show the following important
facts. First, the domain of $\theta$, in which $\tca_{q}>0$,
essentially increases with $q>1$. Hence, validity of Eq.
(\ref{comb5}) with some probabilistic model is not sufficient for
its noncontextuality. Second, measurement statistics of the
experiment with some fixed choice of $\theta$ does violate Eq.
(\ref{comb5}) for one values of $q$ and does not for other ones,
including the standard case $q=1$. For instance, with
$\theta=0.4765$ the inequality (\ref{comb5}) is actually violated
for $1.13<q$ and is not violated with $1\leq{q}\leq1.13$. In other
words, the $q$-inequalities with properly chosen values of $q$
right detect the contextuality of some probability distributions
that cannot be detected by Eq. (\ref{ca151}) with the standard
entropies. Thus, the family of $q$-entropic inequalities provides
much more sensitive criteria for the contextuality. In the same
experimental setup, therefore, we could test violation of the
entire family of $q$-entropic inequalities of the Bell type. The
obtained results can be regarded as an extension and development
of theoretical findings of Refs. \cite{rchtf12,krk12}.

We now consider $q$-entropic inequalities in the more realistic
cases with detector inefficiencies. The writers of Ref.
\cite{rchtf12} considered these cases for
inequalities with the Shannon entropies. It is convenient to
rewrite Eq. (\ref{comb5}) without conditional entropies. Using Eq.
(\ref{chrl}), we have
$H_{q}(X_{j}|X_{j+1})=H_{q}(X_{j},X_{j+1})-H_{q}(X_{j+1})$. Then
the formula (\ref{comb5}) gives
\begin{equation}
0\leq\sum_{j=1}^{4}{H}_{q}(X_{j},X_{j+1})
-H_{q}(X_{1},X_{5})-H_{q}(X_{2})-H_{q}(X_{3})-H_{q}(X_{4})=-{\,}\ca_{q}
\ . \label{ac015}
\end{equation}
Due to detector inefficiencies, we obtain somewhat altered
probability distributions. Hence, calculated entropies will
somehow differ from the entropies involved in Eq. (\ref{ac015}).
The inequality (\ref{ac015}) itself pertains to the
inefficiency-free case, when $\eta=1$. In the single-detector
model, probabilities are given by Eqs. (\ref{pretx}) and
(\ref{pret0}) for the two-observable distribution and by Eq.
(\ref{pret0m}) for the single-observable distribution. By
$H_{q}^{(\eta)}(X_{j},X_{j+1})$ and $H_{q}^{(\eta)}(X_{j})$, we
denote the actual $q$-entropies calculated with such
distributions. If the inequality (\ref{ac015}) is valid, then the
actual entropies satisfy the same formula, namely
\begin{equation}
0\leq\sum_{j=1}^{4}{H}_{q}^{(\eta)}(X_{j},X_{j+1})
-H_{q}^{(\eta)}(X_{1},X_{5})-H_{q}^{(\eta)}(X_{2})-H_{q}^{(\eta)}(X_{3})-H_{q}^{(\eta)}(X_{4})
\ . \label{etc015}
\end{equation}
Indeed, from Eq. (\ref{rlem0}) we immediately write
\begin{align}
H_{q}^{(\eta)}(X_{j},X_{j+1})&=\eta^{q}H_{q}(X_{j},X_{j+1})+h_{q}(\eta)
\ , \label{clem0a}\\
H_{q}^{(\eta)}(X_{j})&=\eta^{q}H_{q}(X_{j})+h_{q}(\eta)
\   . \label{clem0b}
\end{align}
Substituting these expressions, the inequality (\ref{etc015}) is
recast as Eq. (\ref{ac015}) multiplied by factor $\eta^{q}$. In
the single-detector model, therefore, the noncontextuality
hypothesis leads to the family of $q$-entropic inequalities of the
form (\ref{etc015}) with $q\geq1$. The following points should be
emphasized. First, in the considered model violations of Eq.
(\ref{etc015}) are irrelevant to the detection efficiency
$\eta>0$. Second, for fixed $q$ the maximal violation takes place
in the same cases, for which the inefficiency-free inequality
is maximally violated. For the observables
(\ref{123a})--(\ref{45a}) and the state (\ref{meps}), some cases
of the maximal violation were given above in Table \ref{tab1}. In
this regard, $q$-entropic inequalities of the Bell type succeed
properties of more usual inequalities in terms of the Shannon
entropies.

In the second model of detector inefficiencies, probabilities of
the two-observable distribution are expressed by Eqs.
(\ref{pretxx}), (\ref{pretx0}), and (\ref{pret00}). These
two-observable distributions also marginalize to the
single-observable distributions of the form (\ref{pret0m}). By
$H_{q}^{(\eta\eta)}(X_{j},X_{j+1})$ and
$H_{q}^{(\eta\eta)}(X_{j})$, we denote the actual $q$-entropies in
the considered model of inefficiencies. Using Eq. (\ref{rlem00}),
we obtain
\begin{align}
H_{q}^{(\eta\eta)}(X_{j},X_{j+1})&=\eta^{2q}H_{q}(X_{j},X_{j+1})+
\eta^{q}(1-\eta)^{q}\bigl(H_{q}(X_{j})+H_{q}(X_{j+1})\bigr)
\nonumber\\
&+\bigl(\eta^{q}+(1-\eta)^{q}+1\bigr)h_{q}(\eta)
\ . \label{clem1}
\end{align}
On the other hand, the entropy $H_{q}^{(\eta\eta)}(X_{j})$ is
equal to the right-hand side of Eq. (\ref{clem0b}). For brevity,
we introduce the quantity
\begin{equation}
\ca_{q}^{(\eta\eta)}:=-\sum_{j=1}^{4}{H}_{q}^{(\eta\eta)}(X_{j},X_{j+1})
+H_{q}^{(\eta\eta)}(X_{1},X_{5})+H_{q}^{(\eta\eta)}(X_{2})+H_{q}^{(\eta\eta)}(X_{3})+H_{q}^{(\eta\eta)}(X_{4})
\ . \label{caet15}
\end{equation}
In the inefficiency-free case, when $\eta=1$, this term coincides
with the characteristic quantity (\ref{ca15}). Using Eqs.
(\ref{clem0b}) and (\ref{clem1}), we represent the right-hand side
of Eq. (\ref{caet15}) as
\begin{align}
\ca_{q}^{(\eta\eta)} &= \eta^{2q}{\,}\ca_{q}-\Delta_{q}(\eta)
\ , \label{cta5c}\\
\Delta_{q}(\eta) &=
\eta^{q}\bigl(\eta^{q}+2(1-\eta)^{q}-1\bigr)\bigl(H_{q}(X_{2})+H_{q}(X_{3})+H_{q}(X_{4})\bigr)
\nonumber\\
 &+3\bigl(\eta^{q}+(1-\eta)^{q}\bigr)h_{q}(\eta)
\ . \label{cta5d}
\end{align}
The second summand in the right-hand side of Eq. (\ref{cta5d}) is
positive. For $q>1$, the factor $\eta^{q}+2(1-\eta)^{q}-1$ is
negative for some values of $\eta$ near $1$ from below. So, the
first summand in the right-hand side of Eq. (\ref{cta5d}) can take
positive or negative values. The noncontextuality inequality
(\ref{ac015}) implies $\ca_{q}\leq0$. Using measurement
statistics, however, we actually deal with the quantity
(\ref{caet15}). Suppose that measurement data have lead to the
result $\ca_{q}^{(\eta\eta)}>0$. Generally, one cannot conclude
$\ca_{q}>0$ without the following. We must confide that
the violating term $\eta^{2q}{\,}\ca_{q}$ is sufficiently large in
comparison with the additional term (\ref{cta5d}). To compare
these terms, we introduce their ratio
\begin{equation}
r_{q}(\eta):=\eta^{-2q}{\,}\ca_{q}^{-1}\bigl|\Delta_{q}(\eta)\bigr|
\ , \label{rqet}
\end{equation}
which is related to the case $\ca_{q}>0$. To obtain concrete
estimates of $\eta$, we have found numerically the ratio
(\ref{rqet}) in the cases of maximal violation, which are shown in
Table \ref{tab1}. In these cases, the additional term
(\ref{cta5d}) turns to be nonnegative for all $\eta\in[0;1]$. Then
the experimental result $\ca_{q}^{(\eta\eta)}>0$ would witness
$\ca_{q}>0$, i.e. quantum violations of the noncontextuality
hypothesis. However, large values of $\Delta_{q}(\eta)$ can
prevent this, even if the theoretical violation is maximal.
Therefore, used detection schemes should provide the ratio
(\ref{rqet}) to be sufficiently small.

We have calculated $r_{q}(\eta)$ versus $\eta$ for all the cases
listed in Table \ref{tab1}. With respect to $\eta$, we especially
focus an attention on values, which are very close to $1$ from
below. As calculations show, for fixed $q$ the ratio $r_{q}(\eta)$
decreases with such $\eta$ almost linearly, up to the
inefficiency-free value $r_{q}(1)=0$. Due to almost linear
dependence, we can describe each case by the value of Eq.
(\ref{rqet}) for some suitably chosen $\eta$, say, for
$\eta=0.99$. For estimation purposes, one then writes approximate
formula
\begin{equation}
r_{q}(\eta)\approx 10^{2}{\,}r_{q}(0.99){\,}(1-\eta)
\ , \label{linrq}
\end{equation}
which is appropriate within a range of linear behavior. In Table
\ref{tab2}, the value $r_{q}(0.99)$ is presented for the cases of
maximal violation, which are given above in Table \ref{tab1}.
Initially, this value significantly decreases with $q>1$. Further,
it becomes increasing for sufficiently large $q$. In general, the
required detection efficiency is very high. This conclusion
concurs with the efficiency $\eta\approx0.995$, which was claimed
in Ref. \cite{rchtf12} for relations with the Shannon entropies. A
novel point is that, for given $\eta$, the ratio (\ref{rqet})
essentially depends also on $q$. Among $q$-entropic inequalities
for the KCBS scenario with observables (\ref{123a})--(\ref{45a}),
the choice $q=2$ can be recognized as very appropriate. First, the
value of $\max\ca_{q}$ for $q=2$ is almost maximal in comparison
with other (see Table \ref{tab1}). Second, the ratio (\ref{rqet})
in the second model of detection inefficiencies is sufficiently
small for $\eta>0.99$ (see Table \ref{tab2}). Third, properties of
the $q$-entropies are mathematically simpler just in the case
$q=2$. Some of these properties were considered in Ref.
\cite{vajda68}. With the family of $q$-entropic inequalities,
therefore, we can obtain new possibilities for analyzing
measurement data with detection insufficiencies.

\begin{table}
\begin{center}
\caption{\label{tab2}The values of the ratio (\ref{rqet}) for
$\eta=0.99$ and several $q$ in some cases of maximal violation.}
\vskip0.1cm
\begin{tabular}{|c|c|c|c|c|c|c|c|c|c|c|c|c|}
\hline
$q$ & 1.0 & 1.1 & 1.2 & 1.4 & 1.6 & 1.8 & 2.0 & 2.5 & 3.0 & 5.0 & 8.0 & 11.0 \\
\hline
$r_{q}(0.99)$ & 2.0400 & 1.8593 & 1.3065 & 0.7764 & 0.5461 & 0.4294 & 0.3641 & 0.2944 & 0.2793 & 0.3845 & 0.6227 & 0.8656 \\
\hline
\end{tabular}
\end{center}
\end{table}

\section{Conclusions}\label{sec6}

In the paper, we have expressed Bell's theorem in terms
of the conditional $q$-entropies of order $q\geq1$. Formally, the
presented inequalities are based on several useful properties of
the conditional $q$-entropy. One of them is the well-known chain
rule. Other required properties are proved as Lemmas \ref{lem0}
and \ref{lem1}. The latter result is combined with the chain rule
in deriving new $q$-entropic inequalities of Bell's type. The
statement of Lemma \ref{lem0} is used to study the more realistic
cases with detection inefficiencies. The result of Lemma \ref{lem1}
holds for $q\geq1$ and generalizes analogous property of the
standard conditional entropy. From the physical viewpoint, the
noncontextuality hypothesis is a key ingredient of the derivation.
Assuming the existence of a joint probability distribution for the
outcomes of all observations, we have arrived at a principal
conclusion. Namely, the corresponding conditional $q$-entropies of
order $q\geq1$ should satisfy inequalities of the form
(\ref{rthm}). This claim generalizes the previous entropic
formulations of Bell's theorem. In particular, the inequality
(\ref{comb3}) is a $q$-parametric extension of the
Braunstein--Caves inequality \cite{BC88}. Thus, we have shown that
the noncontextuality hypothesis leads to the entire family of
$q$-entropic inequalities of Bell's type. It turns out that these
inequalities are incompatible with the predictions of quantum
mechanics for many values of the parameters.

With the standard conditional entropy, violations of entropic Bell
inequalities were examined for the CHSH scenario in Refs.
\cite{BC88,rchtf12} and for the KCBS scenario in Refs.
\cite{rchtf12,krk12}. We have explicitly considered violations of
the $q$-entropic inequalities in both the scenarios. The following
principal conclusions can be made. First, the derived $q$-entropic
inequalities allow to expand significantly a class of probability
distributions, for which the nonlocality or contextuality are
testable in this way. Using the $q$-entropic inequalities is an
alternative to the approach  with adding some shared randomness
\cite{rch13}. Second, the $q$-entropic inequalities are expedient
in analyzing cases with detection inefficiencies. In the
single-detector model, features of the $q$-entropic inequalities
are quite similar to features of usual inequalities in terms of
the Shannon entropies. In the two-detector model, the use of the
$q$-entropic inequalities can allow to reduce an amount of
required detection efficiency. The obtained conclusions for
various values $q\geq1$ could be tested in the experiment. For the
conventional CHSH inequality in terms of average values, quantum
violation is limited by the Tsirel'son bound. It would be
interesting to obtain upper bounds on possible violations of
$q$-entropic inequalities of Bell's type. Due to the role of
entangled states in quantum information processing, theoretical
results of such a kind may also have a practical significance.

\acknowledgments
I am grateful to Dr. Tobias Fritz and Dr. Rafael Chaves for useful correspondence and
to anonymous referee for constructive criticism.

\end{document}